\documentclass[journal=ancac3,manuscript=article]{achemso}

\author{Rui Qin}
\altaffiliation{these authors contributed equally to this work.}
\email{qinrui.phy@outlook.com}

\author{Zi-Yu Chen}
\altaffiliation{these authors contributed equally to this work.}
\email{ziyuch@caep.ac.cn}

\affiliation[caep]
{National Key Laboratory of Shock Wave and Detonation Physics, Institute of Fluid Physics, China Academy of Engineering Physics, Mianyang 621999, China}

\title{Strain-controlled high harmonic generation with Dirac fermions in silicene}


\begin{document}

\begin{abstract}
Two-dimensional (2D) materials with zero band gap exhibit remarkable electronic properties with wide tunability. High harmonic generation (HHG) in such materials offers unique platforms to develop novel optoelectronic devices at nanoscale, as well as to investigate strong-field and ultrafast nonlinear behaviour of massless Dirac fermions. However, control of HHG by modulating electronic structure of materials remains largerly unexplored to date. Here we report controllable HHG by tuning the electronic structures via mechanical engineering. Using an \textit{ab initio} approach based on time-dependent density-functional theory (TDDFT), we show that the HHG process is sensitive to the modulation of band structures of monolayer silicene while preserving the Dirac cones under biaxial and uniaxial strains, which can lead to significant enhancement of harmonic intensity up to an order of magnitude. With the additional advantage of silicene in compatibility and integration into the current silicon-based electronic industry, this study may open a new avenue to develop efficient solid-state optoelectronic nano-devices, and provide a valuable tool to understand the strong-field and mechanically induced ultrafast nonlinear response of Dirac carriers in 2D materials.

\end{abstract}

\section{Introduction}

The recent active investigations on high harmonic generation (HHG) from solid-state materials has opened up exciting opportunities in both fundamental physics and potential applications\cite{Ghimire2011,Schubert2014,Vampa2015a,Hohenleutner2015,Ndabashimiye2016}. It provides an unique platform for the study of strong-field and ultrafast electron dynamics in the condensed phase\cite{Kruchinin2018}. HHG has been demonstrated as an attracting tool to explore the electronic structure of bulk crystals\cite{Vampa2014,Otobe2016,Osika2017,You2017,TD2017a}. Experimental reports such as all-optical reconstruction of band structure of ZnO\cite{Vampa2015b}, retrieving energy dispersion profile of the lowest conduction band of SiO$_{2}$\cite{Luu2015}, and measurement of the non-vanishing Berry curvature in symmetry-broken $\alpha$-quartz\cite{Luu2018}, etc., mark some of the impressive advancement. HHG in solids also offers a promising approach for the development of novel compact coherent light sources in the extreme ultraviolet (XUV) and soft x-ray spectral region\cite{Garg2018}, as well as for multi-petahertz-frequency optoelectronic\cite{Garg2016} and attosecond photonics\cite{Hammond2017}. High harmonics and/or attosecond pulses can be shaped in terms of polarization and carrier-envelop phase\cite{TD2017b,Langer2017}, and enhanced by means of tailored semiconductors\cite{Sivis2017} and metallic nanostructures\cite{Han2016}. 

In addition to bulk solids, two-dimensional (2D) materials have also attracted considerable attention for HHG which exhibit distinctive electronic properties. Monolayer \textit{h}-BN driven by out-of-plane laser fields has been calculated to exhibit atomic-like HHG while with a more favorable wavelength scaling\cite{TD2018}. Driven by in-plane fields, enhanced HHG efficiency from isolated monolayer MoS$_{2}$ compared to the bulk has been measured\cite{Liu2017}. Besides, there are a few theoretical and experimental investigations of HHG from 2D materials with massless Dirac fermions (MDF), though so far limited to graphene\cite{Mikhailov2007,Bowlan2014,AN2014,AN2015,Cox2017,Yoshikawa2017,Taucer2017}. The study of nonperturbative HHG from such zero-gap 2D materials offers new possibilities to access strong-field and ultrafast nonlinear dynamics of MDF\cite{Baudisch2018}.

Silicene, as the 2D form of silicon and the silicon analogue of graphene, also exhibits exceptional electronic properties such as MDF behavior, linear energy dispersion, and a high Fermi velocity\cite{Vogt2012,Feng2012,Gao2012,Gao2013}. 
It also has some advantages comparing with graphene. For example, it has a stronger spin-orbit coupling (SOC) than graphene, which makes it more suitable for experimentally studying phenomena like quantum spin Hall effects\cite{liu2011quantum}. 
It has a buckled structure and a better tunability of the band gap, which is important for nanoelectronic applications\cite{ni2011tunable}. 
In addition, as silicon being the element of IV family close to carbon, silicene has apparent advantage over graphene in compatibility and integration into the current silicon-based electronic industry. Thus, HHG from silicene, which has not been studied before, is highly relevant not only for investigating the remarkable properties of massless Dirac quasiparticles, but also for developing future integrated all-optical solid-state devices at nanoscale. 

Being amenable to extensive nanoengineering technologies, the electronic properties of 2D materials can display very rich and complex phenomenons and wide tunability, which are expected to have significant influence on HHG process. Realization of controllable HHG in 2D Dirac materials by tuning the electronic properties is important for novel device applications and to infer ultrafast dynamics of MDF.
To date, however, HHG control by tuning the electronic properties has remained largely unexplored. Moreover, the reported theoretical investigations of HHG from Dirac materials are all based on simplified models without properly considering the full electronic and real crystal structures. 

In this work, we report, to our knowledge, the first controllable HHG by tuning the electronic structures via mechanical engineering. HHG in silicene monolayer under different strain profiles are investigated using an \textit{ab initio} approach based on the framework of time-dependent density-functional theory (TDDFT)\cite{Runge1984,Leeuwen1998,Castro2004a}. We use the methods established recently by Tancogne-Dejean \textit{et al}\cite{TD2017a,TD2017b,TD2018}. The first-principles simulation results show that the electronic band structures of silicene and the related HHG are sensitive to strains. Strain engineering can be a promising strategy for controlling HHG and studying MDF dynamics in 2D materials.

\section{Results}
\subsection*{Crystal and band structure of silicene}

\begin{figure*}[htbp]
\centering
\includegraphics[width=0.7\textwidth
]{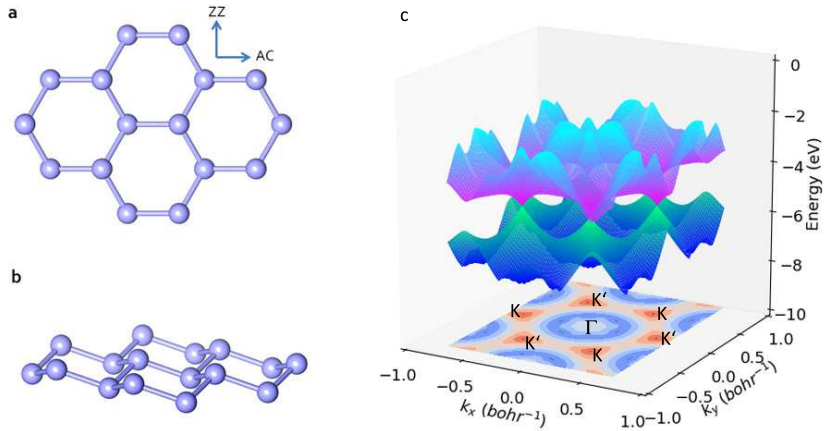}
\caption{\label{structure} (a) Top and (b) side views of the crystal structure of the silicene monolayer. The definition of the armchair (AC) and zigzag (ZZ) directions are shown in panel a. (c) Energy dispersion surface of the highest valence band and lowest conduction band of silicene without strain.  Projection of the highest valence band with high symmetry points marked is also shown.}
\end{figure*}

With a 2D honeycomb lattice structure like graphene, silicene is predicted to favor a low-buckled structure\cite{Cahangirov2009} (Fig. \ref{structure}a, b).  Two special directions of silicene, the armchair (AC) and zigzag (ZZ) directions are marked in Fig. \ref{structure}a. In this study, the geometric structures of silicene are optimized by using the Octopus package. The calculated lattice constant of the hexagonal lattice, Si-Si bond length, and the buckling distance for silicene without strain are 3.81, 2.24, and 0.42 \textup{\AA}, respectively, which are in good agreement with previous calculations\cite{Qin2012,Qin2014}. We also calculate the electronic structure of silicene without strain. The energy dispersion surface clearly shows the Dirac cones at the Brillouin zone (BZ) boundary at the Fermi level (Fig. \ref{structure}c), which induces the massless Dirac fermion behavior near the Fermi level like graphene. Uniform biaxial and uniaxial strain are applied to silicene to investigate the strain effect on HHG. In our calculations, strain $\varepsilon$ is defined as $\varepsilon=(a-a_0)/a_0$ , where $a$ and $a_0$ are the lattice constants with and without strain, respectively. Lattices under the biaxial and uniaxial strains are changed following our previous studies\cite{Qin2012,Qin2014}. The atomic structures are relaxed for the new lattice constant  after each time when we change the lattice constants. For our considered strain, the buckling of silicene decreases with the increasing  strain to release the strain energy. But the buckling will not decrease to zero, and silicene keeps the feature of mixed sp$^3$ and sp$^{2}$ hybridization. Under biaxial strain, the silicene structure maintains the hexagonal symmetry with an elongate lattice constant and chemical bonds. With uniaxial strain applied, the hexagonal symmetry is broken. In this work, we only consider the effect of tensile strain along the ZZ direction for demonstration. Due to Poisson's effect, the lattice constant and chemical bonds along the AC direction is decreased while the lattice and chemical bonds along the ZZ direction is stretched. 

\subsection*{Nonperturbative HHG from silicene}

\begin{figure*}[htbp]
\centering
\includegraphics[width=0.8\textwidth
]{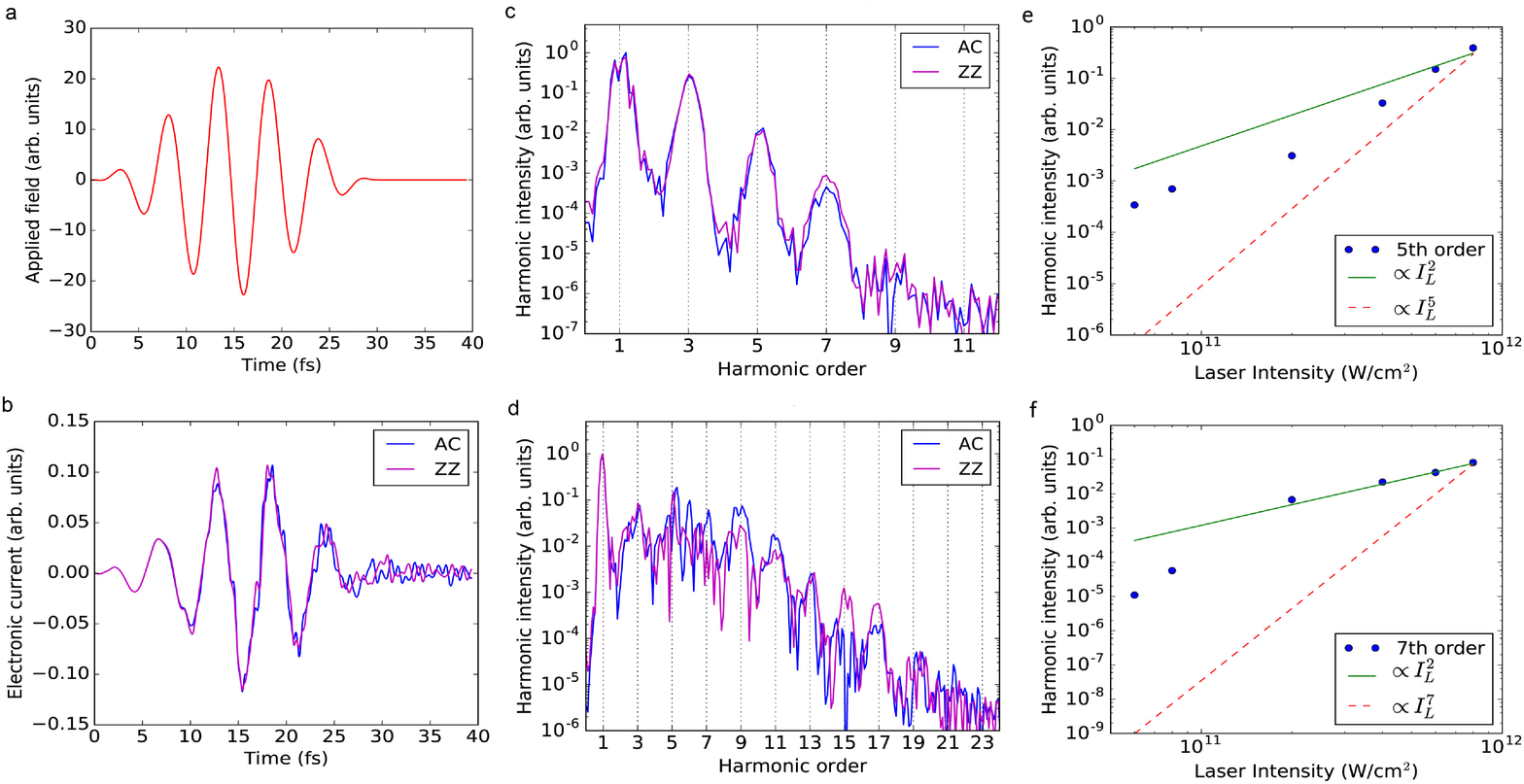}
\caption{\label{no_strain}  (a) The applied vector potential and (b) induced electronic currents with a laser wavelength of 1600 nm, pulse duration of 15 fs, and intensity of $8\times 10^{11}$ W/cm$^2$. Computed HHG spectra from silicene without strain with laser intensity of (c) $8\times 10^{10}$ W/cm$^2$ and (d) $8\times 10^{11}$ W/cm$^2$. Intensity scaling for the (e) 5th and (f) 7th harmonics, demonstrating the harmonic generation in this study is in the nonperturbative regime. The blue dots are the simulation data, while the solid green and dashed red lines are for theoretical fitting.}
\end{figure*}

Figures \ref{no_strain}a-b show the typical profiles of the applied vector potential and induced electronic currents, respectively, where the laser wavelength is 1600 nm and pulse full-width at half-maximum (FWHM) duration is 15 fs. The current waveforms are distorted which contain harmonic components of the driving laser pulse. Figures \ref{no_strain}c shows the HHG spectra from a monolayer silicene without strain with a
driving laser intensity of $8 \times 10^{10}$ W/cm$^2$. Only odd harmonic orders
are present, reflecting the centrosymmetric nature of the crystal lattice. Up to
the 7th harmonic order are obtained. We note that the
experimentally measured HHG from monolayer graphene is up to the 5th order\cite{Taucer2017}
with a driving intensity of $5.7\times 10^{10}$ W/cm$^2$. Considering the
similarity of electronic structures between silicene and graphene, our
simulation results are in good agreement with the experimental observations,
showing the reliability of our simulations. When the laser intensity is increased to $8 \times 10^{11}$ W/cm$^2$, higher-order harmonincs are generated up to the 17th order, as shown in Fig. \ref{no_strain}d. 

Previous studies state that HHG is isotropic in graphene due to rotational
symmetry in the band structure of the Dirac cone\cite{Yoshikawa2017,Taucer2017}. Here we consider laser polarization orienting along two specific directions, i.e., the AC and ZZ directions. Simulation results show evidently different HHG spectra between the two directions, which are pronounced for higher-order harmonics and higher laser intensity (see Figs. \ref{no_strain}c-d). Therefore, even 2D materials with symmetric Dirac cone exhibit anisotropic emission of high harmonics. The anisotropic HHG spectra reflect the asymmetry of energy dispersion in the BZ. Fig. \ref{structure}c shows the band structure is isotropic only in the vicinity of the Dirac points, while it is significantly
anisotropic away from the Dirac points. 

Figures \ref{no_strain}e-f show the intensity of 5th and 7th harmonics as a function of the pump laser  intensity $I_L$, respectively. Both show a characteristic power law scaling of $\propto I_L^2$, demonstrating the HHG in this study is in the nonperturbative regime. Otherwise, they would scale as $\propto I_L^5$ for the 5th-harmonic and $\propto I_L^7$ for the 7th-harmonic if it was in the perturbative limit.

\subsection*{Strain-induced HHG enhancement}

\begin{figure*}[htbp]
\centering
\includegraphics[width=0.7\textwidth
]{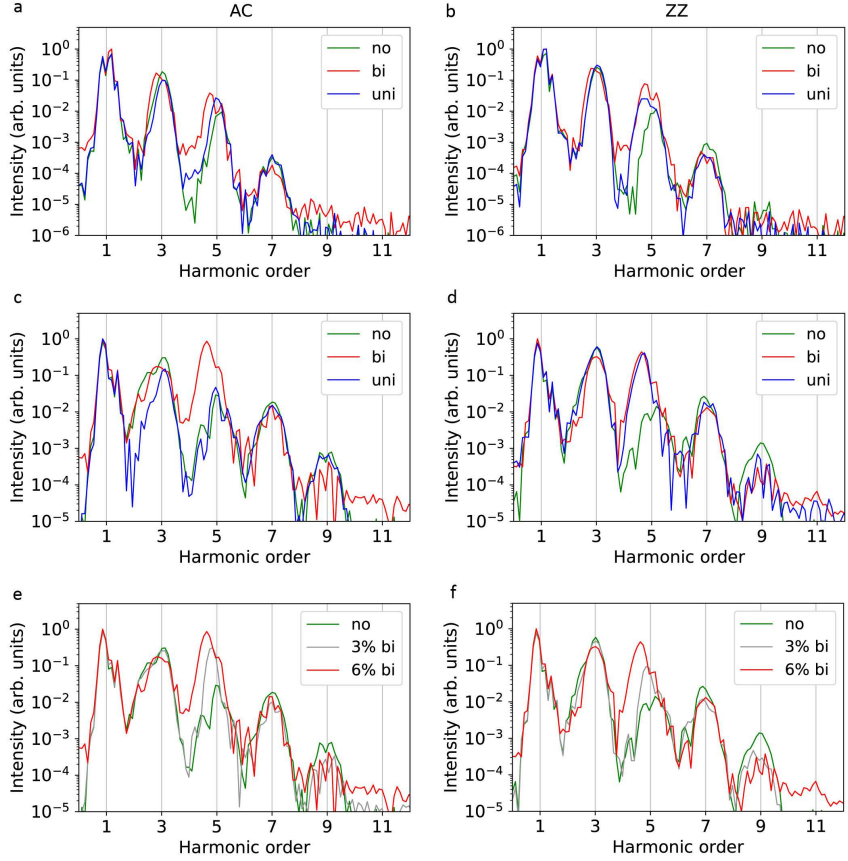}
\caption{\label{strain_control} HHG spectra from silicene under different strain profiles (no: no strain; bi: biaxial strain; uni: uniaxial strain). (a)-(b): laser intensity $I_L = 0.8 \times 10^{11}$ W/cm$^2$ and strain $\epsilon=6\%$. (c)-(d): laser intensity $I_L = 2 \times 10^{11}$ W/cm$^2$ and strain $\epsilon=6\%$. (e)-(f): laser intensity $I_L = 2 \times 10^{11}$ W/cm$^2$ and strain profiles of $\epsilon=0$ (no strain), biaxial strain with $\epsilon=3\%$ and $\epsilon=6\%$.  (a), (c), and (e) correspond to the configurations of laser polarization orienting along the AC directions, while (b), (d), and (f) correspond to ZZ directions.}
\end{figure*}

Fig. \ref{strain_control} shows the computed HHG spectra for silicene under different strain profiles. For a small laser intensity of  $I_L = 0.8 \times 10^{11}$ W/cm$^2$, both biaxial strain and uniaxial strain with $\epsilon=6\%$ can already lead to great enhancement of harmonic intensity compared to the no strain cases, as shown in  Fig. \ref{strain_control}a-b. In addition, the harmonic spectral width is also evidently broadened by applying strains, as can be seen in Fig. \ref{strain_control}b. Increasing the laser intensity to $I_L = 2 \times 10^{11}$ W/cm$^2$, the harmonic intensity is further enhanced. The enhancement can reach an order of magnitude. For the AC and ZZ laser configuration, the 5th harmonic intensity is enhanced by 27 and 29 times  respectively for the 6\% biaxial strain case compared to the strain-free case, as shown clearly in Fig. \ref{strain_control}c-d. The intensity enhancement and spectral broadening also show large anisotropy for laser polarization directed along different directions. For the AC laser polarization configuration, only biaxial strain results in significant increases in harmonic intensity and spectral width; while for the ZZ laser polarization configuration, both biaxial and uniaxial strains lead to harmonic intensity and spectral width enhancement. Fig. \ref{strain_control}e-f show the HHG spectra under no strain ($\epsilon=0$) and biaxial strains with $\epsilon=3\%$ and $\epsilon=6\%$. We see in the strain range considered, the harmonic intensity increases with increasing the strain for both laser polarization configurations, showing the harmonic can be tuned by strain engineering. These results  demonstrate the potential for developing strain-controlled solid-state photonic devices based on HHG in silicene under strain. Apart from these, all the spectra shown in Fig. \ref{strain_control} present odd harmonic orders only, which reflect the inversion symmetry of the crystal are protected even under the biaxial and uniaxial strains.

\subsection*{HHG mechanism from silicene}

\begin{figure*}[htbp]
\centering
\includegraphics[width=0.5\textwidth
]{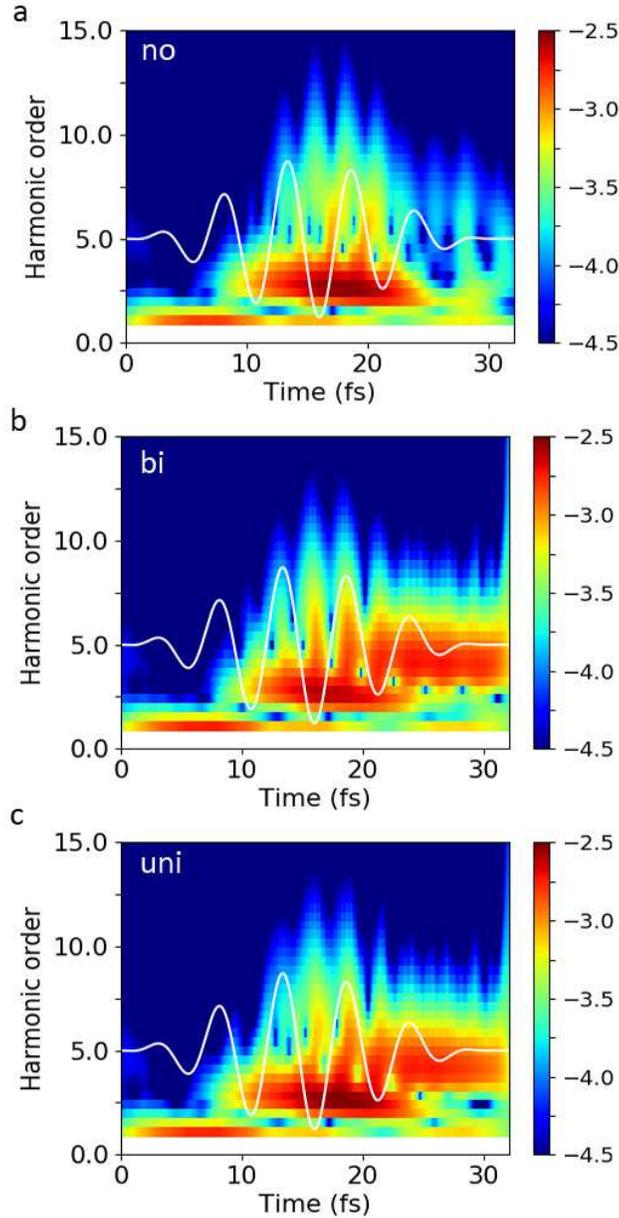}
\caption{\label{wt}  Harmonic order versus time for HHG under different strain profiles: (a) no strain, (b) 6\% biaxial strain; (c) 6\% uniaxial strain. The white curves are the laser waveform profiles. The in-phase feature of harmonic emission as discrete bursts at each peak of the laser field, corresponding to maximum electron acceleration, shows evidence of intraband contribution as the dominate HHG mechanism in this study. The laser polarization orients along the ZZ directions with an intensity of $ 2 \times 10^{11}$ W/cm$^2$. Color shows the spectral intensity (arb. units) on a logarithmic scale.}
\end{figure*}

Nonperturbative HHG in solids originates from two mechanisms, the intraband dynamics, i.e., laser-driven dynamic Bloch oscillation within the non-parabolic band, and interband contribution, i.e., direct electron-hole recombination between the conduction and valence bands. To gain insight into the radiation mechanism, we perform frequency-time analysis of the HHG process. Figure \ref{wt} shows the spectrograms for HHG under different strain profiles (i.e., no strain, biaxial strain, and uniaxial strain), all of which reveal that the high harmonics are emitted as discrete bursts \textit{in phase} at each peak of the laser field corresponding to maximum electron acceleration. The presence of the in-phase feature in the spectrograms instead of recombination trajectories shows strong evidence supporting that intraband contribution is the dominate mechanism\cite{Vampa2017} of the HHG we studied here. It is worth to note that previous model calculations also suggest intraband contribution dominaing the HHG mechanism in monolayer graphene\cite{Taucer2017}.

\subsection*{Strain-induced band structure modulation}
\begin{figure*}[htbp]
\centering
\includegraphics[width=0.65\textwidth
]{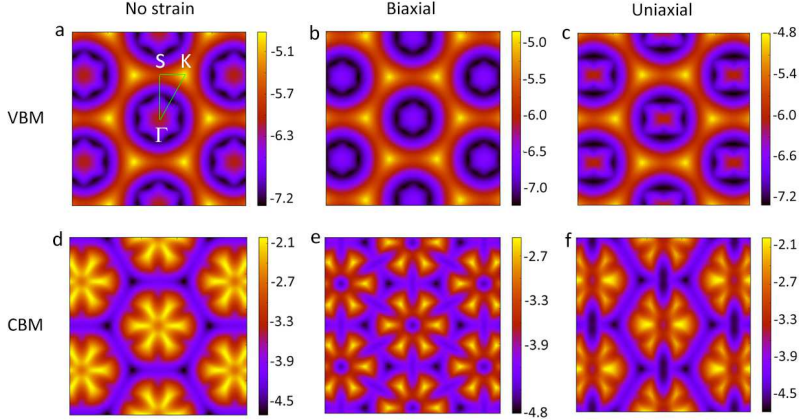}
\caption{\label{2d} Contour plots of energy in reciprocal space of valence band maximum (VBM) ((a)-(c)) and conduction band minimum (CBM) ((d)-(f)) for three different strain profiles: no strain (first column), biaxial strain (second column), and uniaxial strain (third column). High symmetry points $S, K, \Gamma$ in the first Brillouin zone are labeled in panel (a). The units of the color bars are eV.}
\end{figure*}

To understand the HHG enhancement under strains, we now study the electronic structure of silicene under different strain profiles, since the in-plane HHG is closely related to the band structure for 2D materials. First we explore the band structure over the whole BZ. In order to get a simple physical picture, we focus on the highest valence band and lowest conduction band, which are the main contributers of HHG. From Fig. \ref{2d}a-c, we observe that the highest valence band does not change much under various strain profiles comparing to the lowest conduction band counterparts. Therefore we focus on the the lowest conduction band and its variation under strains. Without the presence of strain, the states with low energies all locate at the vicinity of the six K points (i.e., the Dirac cones). The area in which Dirac cones keep the rotational symmetry is also the largest among the three strain profiles (Fig. \ref{2d}d-f). For the biaxial strain case, although the BZ has the same symmetry as the no strain case, the area where the Diac cones have rotational symmetry shrinks. Furthermore, besides the states around the Dirac cones, the states near the $\Gamma$ point shift downward with biaxial strain and finally locate closely above the Fermi level (Fig. \ref{2d}e). For the uniaxial case, the BZ and band structure have a lower symmetry than those of the other two strain cases. Similar with the biaxial case, part of the states shift down towards the Fermi level with uniaxial strain, but they locate near the S point and show high directivity in $\Gamma$-S direction (Fig. \ref{2d}f). The lower the states in the conduction bands are, the more easily the free carriers can be excited by the laser field from the highest valence band to these lowered states, which can result in enhanced HHG.

\begin{figure*}[htbp]
\centering
\includegraphics[width=0.65\textwidth
]{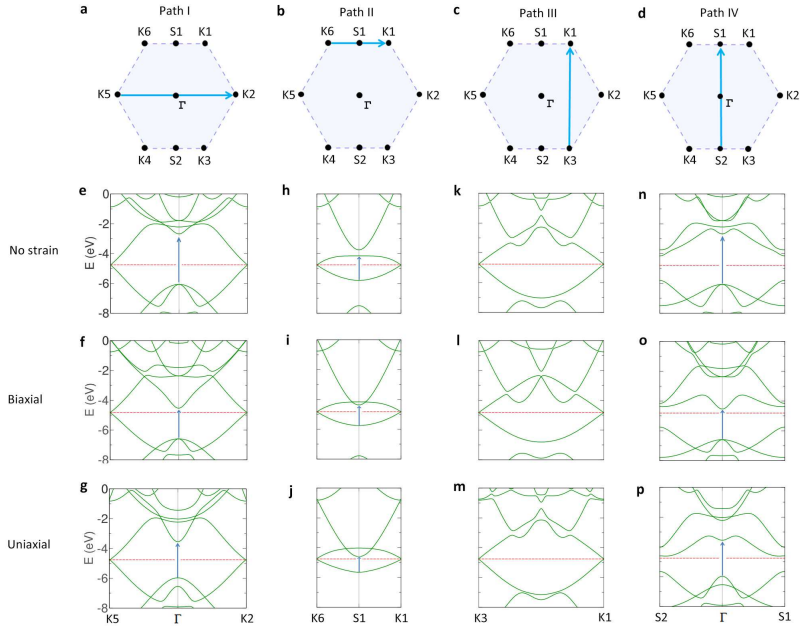}
\caption{\label{bandpath} (a)-(d) Schematics of four representative paths in the Brillouin zone which correspond to the carrier dynamics. Band structures in the reciprocal space along (e)-(g) path I, (h)-(j) path II, (k)-(m) path III, and (n)-(p) path IV. The red dashed lines denote the Fermi level. The first two columns correspond to the configuration of laser polarization along the ZZ direction, and the last two columns correspond to the configuration of laser polarization along the AC direction.}
\end{figure*}

We next analyse the paths in the BZ which correspond to the MDF dynamics in the reciprocal space. For the sake of better illustration, we choose four representative paths where the states change significantly under strain and have major contributions to the variation of HHG process. Path I and II correspond to the ZZ laser polarization configuration (Fig. \ref{bandpath}a-b), while path III and IV to the AC laser polarization configuration (Fig. \ref{bandpath}c-d). Notice the direction in reciprocal and real space are opposite. Under the uniaxial strain, the Dirac points will shift parallel to the $\Gamma-$K2 direction\cite{Qin2014}. However, they only shift very slightly in the considered strain of 6\%, thus we use the same paths for the uniaxial strain case as the other two strain cases for consistency. 

We first investigate MDF dynamics in the ZZ laser polarization configuration. Considering path I, when there is no strain, most of the states near the Fermi level are from the $\pi^*$-band. Besides, there is only a small overlap between the $\pi^*$- and $\sigma^*$-band near the $\Gamma$ point in conduction band minimum (CBM) (Fig. \ref{bandpath}e). With the presence of strain, the overlap between the $\pi^*$- and $\sigma^*$-band becomes larger, and the states near the $\Gamma$ points shift towards the Fermi level (Fig. \ref{bandpath}f-g). The hybridization of $\pi^*$- and $\sigma^*$-band leads to a larger anharmonicity of CBM comparing to the no strain case, which could affect the HHG process via the intraband mechanism. Also the states near the $\Gamma$ point in the biaxial strain case shifts larger than that in the uniaxial strain case and get close to the Fermi level, which will influence the population in the CBM and even higher conduction bands. In path II, we also observe this hybridization and enhanced anharmonicity of the CBM under strains (Fig. \ref{bandpath}h-j). The states near the S point shift downward under both strains, and they get slightly lower in the uniaxial case than that in the biaxial case.

For the case of AC laser polarization configuration, from the band structures in path III, a small overlap between the $\pi^*$-
and $\sigma^*$-band can be observed in the biaxial strain case (Fig. \ref{bandpath}l) while these two bands remain untouched in the no strain (Fig. \ref{bandpath}k) and uniaxial strain (Fig. \ref{bandpath}m) cases.  Aslo, part of the CBM shift downward slightly in the biaxial strain case while change little in the unaxial strain case. For all strain cases, most of the low-lying states in Path III still locate near the Dirac points (Fig. \ref{bandpath}k-m). In contrast, part of CBM shift downward to the Fermi level in path IV under both biaxial and uniaxial strains, and new states emerge close to the Fermi level, which locate near the $\Gamma$ and S points for the biaxial (Fig. \ref{bandpath}o) and uniaxial (Fig. \ref{bandpath}p) strains, respectively. In addition, states near the $\Gamma$ point in the biaxial case shift much larger than states near the S point in the uniaxial case, which can lead to a larger electron population change.

\subsection*{Strain-induced population difference}

\begin{figure*}[htbp]
\centering
\includegraphics[width=0.65\textwidth
]{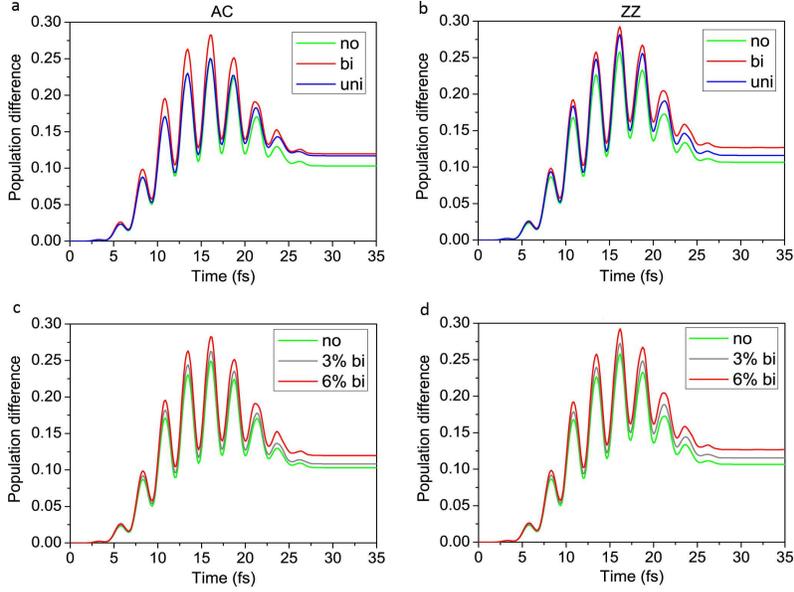}
\caption{\label{pop}  Temporal evolution of the population difference in conduction bands of silicene for different strain profiles. (a)-(b): comparison of population difference for cases of no strain (green), 6\% biaxial strain (red), and 6\% uniaxial strain (blue). (c)-(d): comparison of population difference for cases of no strain (green), 3\% biaxial strain (gray), and 6\% biaxial strain (red). Panels (a) and (c) are for the AC laser polarization configuration, while (b) and (d) are for the ZZ configuration. For all the panels, the laser intensity is $I_L = 2 \times 10^{11}$ W/cm$^2$. 
Here the population difference is defined as $|N(t)-N(t=0)|$ ,  where $N(t)$ and $N(t=0)$ denote the electron population at time \textit{t} and ground state (\textit{t}=0). The population are summed over the three lowest conduction bands, since the contribution of electron population in even higher conduction bands are quite small and negligible.}
\end{figure*}

The above analyses of strain-induced band structure modulation imply increased number of excited electrons into conduction bands, which can affect the HHG process and result in HHG enhancement. We then inspect the time evolution of electron population difference in various bands for the strain-free and different strain cases to investigate this effect (see the Methods section for more technical details of population calculation). 
Here we compare the total electron population of the three lowest conduction bands for different strain profiles, since the contribution of electron population in even higher conduction bands are quite small and negligible. 
Figure \ref{pop} shows the electron population evolution with a laser intensity of $I_L = 2 \times 10^{11}$ W/cm$^2$. For the AC laser polarization configuration, the biaxial strain case shows the largest electron population while the population for the strain-free and uniaxial strain cases are similar.  For the ZZ laser polarization configuration, the electron population is larger in the biaxial and uniaxial strain cases than that in the no strain case. 
Besides, we calculate the electron population under a intermediate biaxial strain of 3\%. As shown in \ref{pop}c-d, the electron population increases with increasing the strain from $\epsilon=0$ to $\epsilon=6\%$. These strain-induced population difference are in good agreement with the corresponding harmonic spectral profiles shown in Fig. \ref{strain_control}c-f, supporting our band structure analysis and suggesting the enhanced carrier population via strain-induced band structure modulation may be responsible for the harmonic enhancement under strains. 
As mentioned above, intraband contribution dominates the HHG process in this study. However, we emphasise that HHG would be enhanced for both intraband and interband mechanisms if  the population was increased. Other than the population increase, it should also be noted that the strain-induced band structure modulation also changes the curvature of the energy dispersion, which could be another possible contribution to the changed HHG intensity via the intraband mechanism.
Apart from these, the presence of strain changes the lattice structures of silicene, which changes the charge density. Moreover, strain also brings profound modulation to the electronic band structures, and consequently, the carrier density is further varied. These strain effects can lead to modification of the nonlinear refraction index of the material and thus phase changes of the light waves.  Various nonlinear processes such as self-phase modulation and cross phase modulation may contribute to the spectral broadening shown in Fig. \ref{strain_control}.

\section{Conclusions}
In summary, we have investigated high-harmonic emission from monolayer silicene driven by in-plane electric fields using first-principles TDDFT calculations. We show that HHG in silicene can be actively controlled in terms of intensity and spectral width by modulating the electronic properties of the 2D material via strain engineering. For the parameter regime we considered, HHG intensity can be enhanced by an order of magnitude by applying biaxial and uniaxial strain of 6\%. We identify intraband contribution is the dominate HHG machanism. The strain-induced band structure modulation can lead to increased number of excited electrons, which eventually results in enhanced HHG. Strain engineering of HHG should be applicable in other similar 2D materials with MDF, e.g., graphene. Many other approaches, such as using substrate, doping, and electrostatic gating, can also be introduced to tune the electronic properties and HHG in 2D materials.  It is worth noting that substrates are generally needed in lab or applications, which usually should simultaneously fulfil the requirements of large applied strain and weak coupling. Polymers, such as PmPV and PDMS, may be considered as potential substrates for these requirments, because they  are both expandable and interact weakly with silicene\cite{Chen2015}.  This study opens a new avenue to develop novel solid-state optoelectronic devices at nanoscale. It can also provide important insights for the investigation of strong-field and mechanically induced nonlinear response of Dirac carriers in 2D materials. By measuring the time evolution of HHG, it should be possible to obtain ultrafast dynamical information of the Dirac fermions. Besides, HHG offers a valuable tool to understand the strain effect on silicene that is important for the 2D material growth.

\section{Methods}

Silicene structures are studied by using the semiperiodic supercell model. Silicene structures without strain and under biaxial and uniaxial strains are modelled by the hexagonal primitive cell containing two silicon atoms. A vacuum of 30 Bohr, which includes 3 Bohr of absorbing regions on each side of the monolayer, is chosen to eliminate the interactions between adjacent silicene sheets and avoid the reflection error in the spectral region of interest. All calculations are performed by the Octopus package\cite{Andrade2015,Castro2006,Andrade2012}. 
SOC is expected to have little influence on MDF dynamics and HHG under strong field in silicene, since the band structure only changes slightly with SOC \cite{matthes2013massive}and the SOC gap ($\sim$1.55-2 meV) \cite{liu2011quantum,matthes2013massive} is much smaller than the applied photon energy. Thus SOC is not considered in this work. 
The ground state electronic structure properties and geometric structure relaxation are performed within the density functional theory (DFT) framework in the local density approximation (LDA)\cite{Marques2012}. Time evolution of the wave functions and time-dependent electronic current are studied by propagating the Kohn-Sham equations in real time and real space\cite{Castro2004b} within the time-dependent DFT (TDDFT) framework in the adiabatic LDA (ALDA). The real-space spacing is 0.4 Bohr. A $80\times 80\times 1$ Monkhorst-Pack \textit{k}-point mesh for the BZ sampling is used. The fully relativistic Hartwigsen, Goedecker, and Hutter (HGH) pseudopotentials are used in all our calculations. 

The laser is described in the velocity gauge. A sin-squared envelope and a carrier-envelope phase of  $\Phi= 0$ is used for the laser profile. The laser wavelength is $\lambda=$1600 nm (corresponding to a photon energy of 0.77 eV) and FWHM pulse duration is $\tau$=15 fs. The linearly polarized laser field is normally incident onto the silicene sample so that the driving electric field is in the plane of the monolayer. HHG in silicene are shown to exhibit nonperturbative characteristic in the considered laser intensity range of $I_0=0.8\times10^{11}$ W/cm$^{2}$ to $I_0=8\times10^{11}$ W/cm$^{2}$.

The HHG spectrum was calculated from the total time-dependent electronic current  $\textbf{j}(\textbf{r},t)$ as:
\begin{equation}
\mathrm{HHG}(\omega) = \Big| \mathrm{FT} \Big(\frac{\partial}{\partial t} \int \textbf{j}(\textbf{r},t) \ \mathrm{d}^3 \textbf{r}  \Big) \Big|^2,
\end{equation}
where FT denotes the Fourier transform. 

Electron population of the band was calculated by the projection of the time-dependent Kohn-Sham states on the the ground-state Kohn-Sham states. As the \textit{n}-th state evolves in time, it has some possibility to transit to other states and thus contain other ground-state components. The population that any occupied state has evolved into the \textit{m}-th ground-state Kohn-Sham band is:
\begin{equation}
P(m,t)=\int_{\mathrm{BZ}}  \sum_{n} f_{n,k} |\langle\phi_{n,k}(t)|\phi_{m,k}(0)\rangle|^2 \mathrm{d} \textbf{k} 
\end{equation}
where $\phi_{n,k}(t)$,$\phi_{m,k}(0)$, $f_{n,k}$, BZ denote the time-dependent Kohn-Sham state at \textit{n}-th band at k-point \textit{k}, the ground-state (t = 0) Kohn-Sham state at \textit{m}-th band at k-point \textit{k}, and the occupation of Kohn-Sham state at \textit{n}-th band at k-point \textit{k}, integration over the whole Brillouin zone, respectively.




\bibliography{refs}

\begin{acknowledgement}
We acknowledge financial support from the National Natural Science Foundation of China (NSFC) (11705185) and the Presidential Fund of China Academy of Engineering Physics (CAEP) (YZJJLX2017002). 
\end{acknowledgement}



\end{document}